\documentclass[12pt,preprint]{aastex}

\begin{document}


\title{Spectroscopic Observations of Hot Lines Constraining Coronal
Heating in Solar Active Regions}
\author{S. Patsourakos\altaffilmark{1,2}}

\author{J. A. Klimchuk \altaffilmark{3}}

\altaffiltext{1}{ Naval Research Laboratory, Space Science Division,
Washington,  DC 20375}
\altaffiltext{2}{Center for Earth Observing and Space Research, Institute for Computational Sciences - College of Science,
George Mason University, Fairfax, VA 22030}
\altaffiltext{3}{ NASA Goddard Space Flight Center, Code 671,
Greenbelt, MD  20771}



\begin{abstract}
EUV observations of warm coronal loops suggest
that they are bundles of unresolved strands that
are heated impulsively to high temperatures
by nanoflares. The plasma would then have the observed properties
(e.g., excess density compared to static equilibrium) when it cools into
the 1-2 MK range. If this interpretation is correct, then very hot emission
should be present outside of proper flares. It is predicted to be vey faint,
however. A critical element for proving or refuting this hypothesis is the existence
of hot, very faint plasmas which should be at amounts predicted by impulsive heating.
We report  on the
first comprehensive spectroscopic study of hot plasmas in active regions. Data
from the EIS spectrometer on Hinode were used to construct emission measure
distributions in quiescent active regions in the 1-5 MK temperature range. The
distributions are flat or slowly increasing up to approximately 3 MK and then
fall off rapidly at higher temperatures. We show that active region models based
on impulsive heating can reproduce the observed EM distributions relatively
well. Our results provide strong new evidence that coronal heating is impulsive
in nature.
\end{abstract}




\section{Introduction}
Almost 10 years ago it was realized that warm ($\approx$ 1 MK)
active region (AR) loops seen in the EUV cannot be explained by
static equilibrium theory. For example the observations showed that
these loops are appreciably more dense than what static loop models
would predict (e.g.,  Aschwanden, Schrijver $\&$ Alexander 2001;
Winebarger, Warren $\&$ Mariska 2003; Warren, Feldman $\&$ Brown
2008; Warren et al. 2008).

It was thus proposed that coronal loops could be collections of
unresolved strands which are heated impulsively by small-scale
heating events (i.e. nanoflares) to high temperatures of several MK
which then cool down to the 1 MK range to give rise to the observed
EUV loops (e.g., Cargill 1993; and the review of Klimchuk 2006 and
references therein). This paradigm proved quite successful at
reproducing several key observational aspects of the 1 MK loops
(e.g.  Klimchuk 2006). Note here that impulsively heated
sub-resolution strands can also be invoked to explain the emissions
from areas not containing resolved coronal loops, i.e., diffuse
background areas. The primary difference between  loops and
background could be that for loops, nanoflares occur in a somewhat
coordinated fashion over a distance comparable to the loop diameter,
whereas for the background, they occur at rather random times and
with a broader spatial distribution.

However, the details of the heating process are essentially lost by
the time an impulsively heated loop cools though the 1 MK range
(e.g., Winebarger $\&$ Warren 2004; Patsourakos $\&$ Klimchuk 2005; Parenti et al. 2006). 
Therefore, one has to search for
signatures of the postulated impulsive heating in higher temperature
emissions in order to prove or refute this picture of coronal
heating. Observing in spectral lines is preferable, since they are
formed over a rather narrow temperature range which allows a more
precise and unambiguous temperature determination compared to narrow
and broad band imaging. Observations in hot emissions could then be
viewed  as a  true ``smoking gun'' of the impulsive heating (e.g.,
Cargill 1995; Patsourakos and Klimchuk 2006). One
difficulty with observing the hot spectral emissions is that they
are predicted to be quite faint (e.g., Patsourakos $\&$ Klimchuk
2006;  Bradshaw $\&$ Cargill 2007; Reale $\&$ Orlando 2008;
Klimchuk, Patsourakos, \& Cargill 2008). Moreover, there  exist very
few observations of hot spectral emissions in quiescent active
regions taken with desirable spatial resolution.

With this work we address the following important questions. Is
there any evidence of the elusive hot line emissions? Are there any
spectroscopic signatures of hot ($>$ 3 MK) loops in ARs? And are the
intensities in hot lines consistent with the predictions of
impulsively heated models? It is very timely to address these
questions, given the availability of the new, high-quality
data-stream from the Extreme Ultraviolet Imaging Spectrometer (EIS)
of Hinode. In particular, the larger effective area of EIS compared
to its predecessors is critical  for observing the important yet
faint hot line emissions.

\section{Observations and Data Analysis}
EIS (Culhane et al. 2007;  Korendyke et al. 2006) is an imaging
spectrometer operating  in two ranges of the EUV  (171-212 \AA \,
and 245-291 \AA). These windows contain spectral lines formed in the
range $\approx$ 0.05-15 MK. Our observations were taken from
22:40-23:35 UT on 30 June 2007; EIS study {\it
HPW008\_FULLCCD\_RAST}. 128 steps of the scanning mirror were made,
with 30 s exposures taken at each slit position. The pixel size is 1
$\mathrm{{arcsec}^{2}}$. An area of 128$\times$128
$\mathrm{arcsec^{2}}$ was rastered and full CCD readouts were sent
to Earth. The target was a small and simple active region (NOAA AR
10961) not very far from disk center. The raw data were processed
with the standard {\it eis\_prep} routine which among other things
subtracts the dark current, identifies hot, warm and dusty pixels,
and finally applies the absolute photometric calibration to the
data.

\begin{center}
\begin{figure}
\epsscale{0.8} \plotone{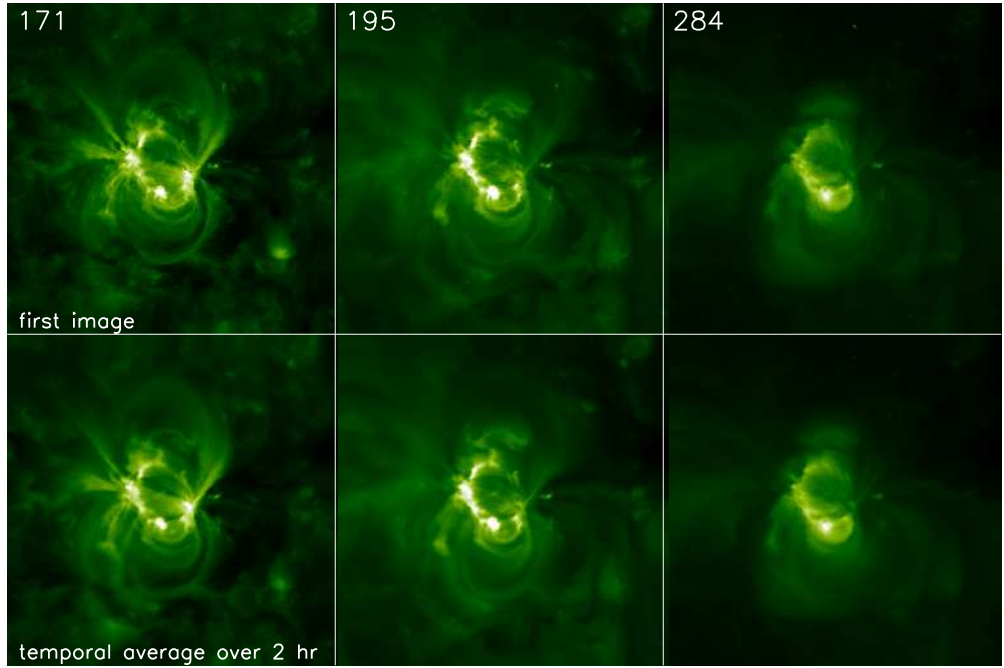}
\caption{Images in the 171, 195 and 284 channels of EUVI of STEREO A
taken around 22:00 UT on 30 June 2007 (upper panel) and temporal
average  of all corresponding images for the same channels for the
time interval of $\approx$ 22-24 UT on 30 June 2007 (lower panel).
The square root of intensity is displayed. Same scaling is applied
to images from the same wavelength. } \label{fig:img_cube}
\end{figure}
\end{center}

The target AR did not evolve substantially during our observations.
Inspection of Hinode XRT (Golub et al. 2007) and STEREO SECCHI/EUVI
(Howard et al. 2007) movies  in SXR and the EUV respectively
revealed a number of distinct loops, most of which were visible
during the entire time of the EIS raster. The upper panels in Figure
\ref{fig:img_cube} show images in the 171, 195 and 284 channels of
EUVI (from the Ahead STEREO spacecraft) taken around 22:00. The
lower panels show corresponding averages of co-aligned images from
the interval 22:00-24:00. The similarity in the appearance of the
instantaneous snapshot and time average for each channel is
indicative of a lack of major evolution.

\begin{center}
\begin{figure}
\epsscale{0.8}
\plotone{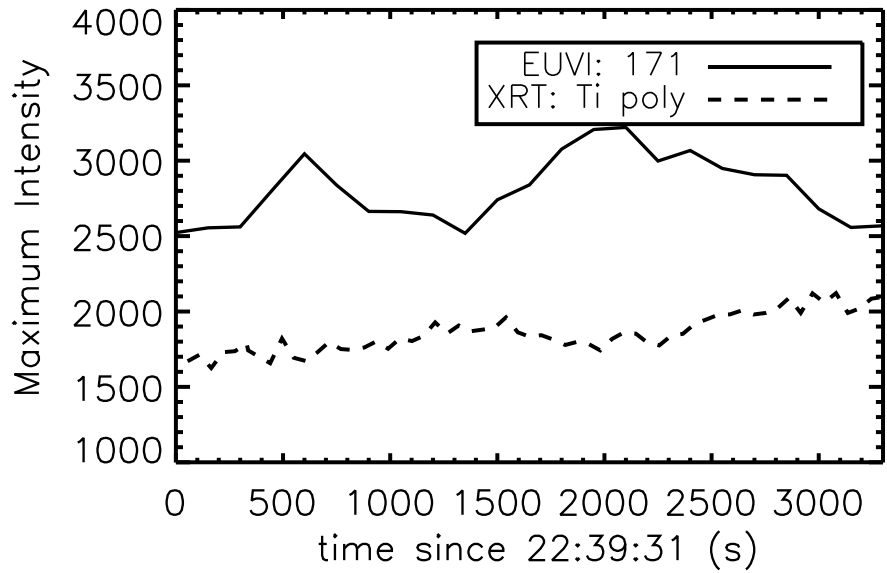}
\caption{Light-curves of the maximum intensities of the active
region as observed in the 171 channel of EUVI on STEREO Ahead
spacecraft (solid) and the Tipoly channel of XRT/Hinode (dashed
line) during the period of EIS observations. Intensities in
arbitrary units.} \label{fig:lc}
\end{figure}
\end{center}

To quantify this, in  Figure \ref{fig:lc} we plot the time evolution
of the maximum intensities in the observed AR for the 171 ($\approx$
1 MK) and Tipoly ($\approx > $ 5 MK)  channels of EUVI and XRT
respectively. We see that the maximum intensities do not vary by
more than $\approx$ 20-30 $\%$ during the 1 hr of our observations.
The variation in the spatially averaged intensities is even less.
Note that the pixel of maximum intensity changes location every 2-4
exposures, an indication that some variability is present.

It is important to stress that the lack of perceived evolution does
not preclude the possibility that dramatic changes are happening on
a subresolution scale.  In particular, if loops and/or the diffuse
background are heated by nanoflares within unresolved strands, then
the strands will be evolving rapidly even when the observed
large-scale structures appear to be steady.

For further analysis we selected a series of well-defined warm and
hot lines formed in the temperature range $\approx$ 1-5 MK (Table
1). We used line identifications proposed by the EIS team (Young et
al. 2007; Brown et al. 2008; Del Zanna 2008; Warren et al. 2008).
Significant effort has has been made to identify ''clean" hot lines
in the EIS wavelength range (Del Zanna 2008; Warren et al. 2008),
and such lines are used in our analysis.

\begin{center}
\begin{figure}
\epsscale{0.8}
\plotone{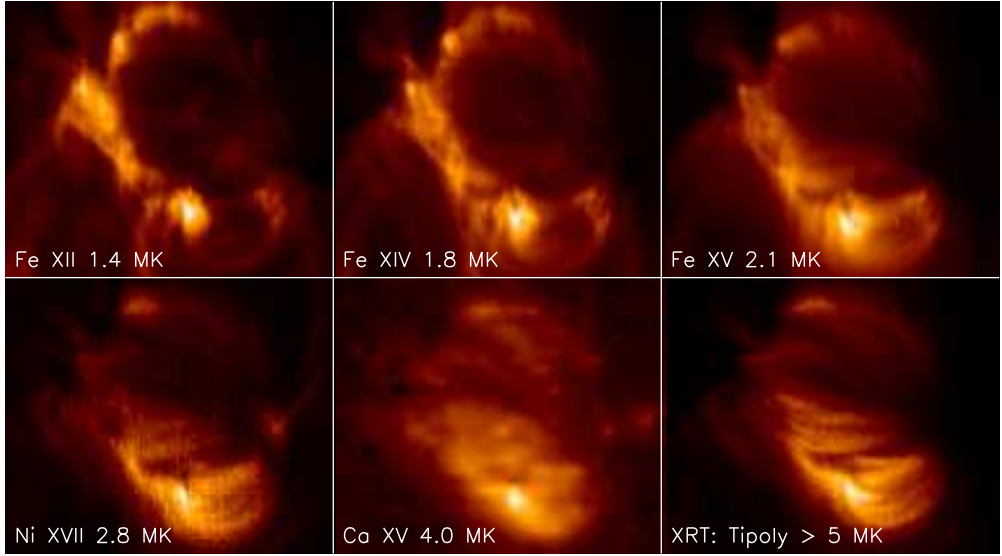}
\caption{Images of the observed AR in 5 spectral lines observed by
EIS and in the Tipoly channel of XRT. Square root of intensity is
displayed, and each image is scaled individually. Formation
temperatures are indicated. The observed field is 128$\times$111
$\mathrm{{arcsec}^{2}}$, and spatial shifts between EIS data
recorded in the 2 different CCDs have been corrected.}
\label{fig:arim}
\end{figure}
\end{center}

Images of the AR in several of the lines are displayed in Figure
\ref{fig:arim}, together with a Tipoly image from XRT. For the
strong and isolated Fe XII, Fe XIV, Fe XV, and Ni XVII lines the
images were constructed by simply integrating the background
subtracted profiles at each pixel. For Ca XV  we first binned the
data in 2$\times$2 pixels to increase the signal-to-noise ratio.
Then, for each macro-pixel we employed a two-Gaussian fit to account
for a nearby Fe XIII line at $\approx 201.12$ \AA.

Several remarks can be  made about morphology of the hot plasma seen
in Figure \ref{fig:arim}.
The emission in Ca XV (4 MK) is
rather widespread in the active region and not just concentrated in
a small number of discrete features. A few rather fuzzy loop
structures are discernable. The overall morphology in this line is
similar to that seen in the broad-band XRT image in the Tipoly
channel which is sensitive to plasmas $\approx >$ 5 MK. Note however
that the XRT image looks sharper than the Ca XV image. This is due
partly to the higher signal-to-noise ratio of XRT data (XRT is a
broad-band instrument while EIS takes monochromatic images over
small wavelength windows) and partly to the higher spatial
resolution of XRT. The Ca XV image also seems smoother because of
the 2$\times$2 pixel binning.

Unfortunately, the signal in the remaining hot lines at single
pixels or macro-pixels of reasonable size (e.g. 2$\times$2,
4$\times$4) is too faint to construct images like we did for the
lines shown in Figure \ref{fig:arim}.  For example, the strongest Fe
XVII line is about 100 times fainter than the  Fe XII line at 195.12
\AA. We therefore computed an average profile over the entire AR for
each of the lines of Table 1. We then fitted the average profile
with one or two Gaussians plus a linear background (a double
Gaussian was used when another line was in the vicinity of the line
of interest). Finally, we determined an average intensity from the
fit.

It was necessary to account for two effects before combining the
profiles to obtain the averages. First, the two wavelength ranges of
EIS are imaged on separate CCD detectors, and there is a relative
offset of $\approx$ 2 and 17 pixels in the x and y directions.  We
restricted our averages to the common parts of the two detectors.
Second, the spectral lines exhibit a quasi-periodic drift that is
believed to be caused by thermal variations of the spacecraft during
its $\approx$ 90 min orbit. The amplitude of
the drift corresponds to a Doppler shift of several 10
$\mathrm{km/s}$. We corrected for this effect in the following
manner (see also Warren et al. 2008). First, we determined the
centroid (first moment) of the strong Fe XII line at 195.12 \AA \,
at each location along the slit. Then, we calculated the average
centroid along the slit for each raster position. Finally, we
interpolated between the slit averages to determine the wavelength
correction as a function of time. This makes the implicit assumption
that  the observed drifts of the spectral lines are not of solar
origin.  We verified the robustness of the above process by: (1)
repeating it for another strong line (Fe XV 284.16 \AA) which is
recorded on the other CCD and is formed at a different temperature;
(2) computing the centroid of the slit-averaged profile at each
raster position, rather than the average centroid of individual
profiles. In both cases, the deviation from the original scheme does
not exceed a few $\mathrm{km/s}$.

As a last step in the analysis, we calculated the emission measure
distribution $EM(T)$ of the AR using the spatially averaged
intensities and the latest version of CHIANTI v5.2 (Dere et al.
1999; Landi et al. 2006). We made the usual assumption that the bulk
of the line emission originates from a temperature interval $\Delta
\log T = 0.3$ that is centered on the temperature of peak formation.
This was shown to be a reasonable assumption even for complex plasma
distributions like those from impulsively heated models (Klimchuk
$\&$ Cargill 2001). Notice that the employed ions (Fe, Ca and Ni)
all have a low First Ionization Potential, and therefore the
relative intensities should be largely independent of the assumed
set of elemental abundances. We chose the abundances of Feldman (1992).
Finally, error bars in the determined $EM$ were calculated by
quadratically combining the errors in the fitted intensities plus a
30$\%$ uncertainty in the photometric calibration of EIS, based on
pre-flight measurements (Culhane et al. 2007).

The $EM$ distribution of AR 10961 is shown in Figure \ref{fig:em}.
The distribution is relatively flat in the temperature range from
$\approx$ 1-4 MK, and then falls off by almost a factor of 10 at our
hottest line at 5.0 MK.  Lines formed at similar temperatures are
generally consistent to within the error bars (vertical lines),
though small discrepancies remain that may be due to uncertainties
in the atomic physics.

To determine whether this distribution is typical of most active
regions, we repeated our analysis on 8 additional datasets obtained
between January and September 2007. Only a subset of the lines in
Table 1 were used: Fe XII, Fe XV, Ni XVII and Fe XVII (254.87
\AA\,).  The results are shown in Figure \ref{fig:ar_em}. In all
cases, the $EM$ distribution is either flat or mildly increasing up
to 3 MK and falls off by 1-1.5 orders of magnitude at 5 MK.
Similar distributions have been reported
from smaller datasets (e.g., Brosius et al. 1996; Watanabe et al.
2008). It is intriguing that the $EM$ of the coolest line, Fe XII
formed near 1.4 MK, varies by only a factor of 2-3 for all of our
examples, whereas the $EM$ of the hottest line, Fe XVII, is much
more variable. The ratio of the two lines ranges from $\approx$
5-50. This suggests that hot emissions are the most sensitive
indicators of the coronal heating mechanism, as indicated above and
discussed shortly.

Before proceeding, we return to the question of whether spatially
averaged intensities are representative of active regions as a
whole, especially for faint hot plasmas. For instance, a transient
brightening and/or a localized bright spot may dominate the
averages. We can rather safely exclude this possibility for the
following reasons. First, the light curves of Figure \ref{fig:lc}
show that no appreciable brightening took place during the
observations of AR 10961. Second, the hot Ca XV emission is widely
distributed throughout the active region (Figure \ref{fig:arim}). We
have confirmed that the conspicuous bright spot in the lower middle
part of the image does not greatly influence the average. Third, we
have performed sit-and-stare observations of this active region in
the even hotter Fe XVII lines (maintaining a single slit position
for $\approx$ 1 hour).  There is significant emission along most of
the slit, with enhancements in places that are known to cross loops
observed in warmer lines like Fe XV.

\begin{center}
\begin{figure}
\epsscale{0.8}
\plotone{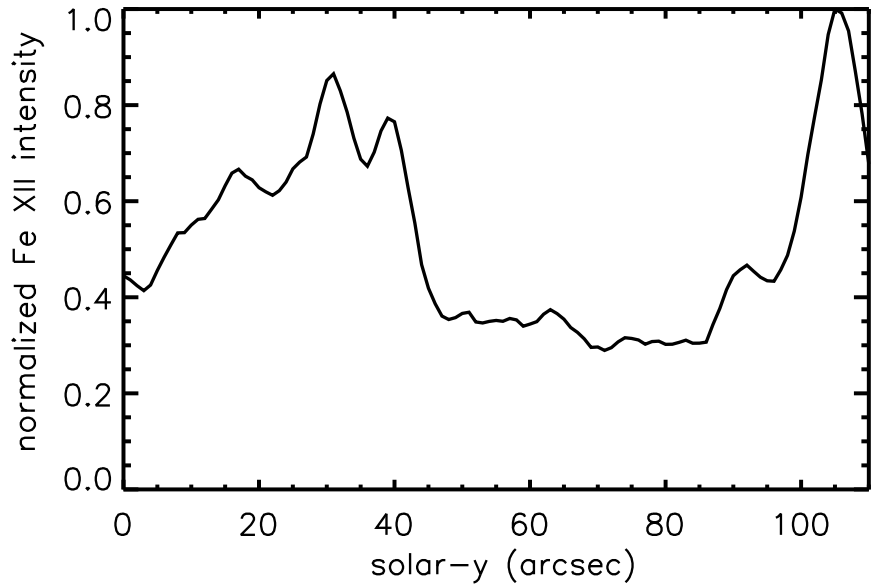}
\caption{Normalized Fe XII 195.12 \AA \, intensities as a function of solar-y for a vertical cut
in the middle of  the Fe XII image of Figure \ref{fig:arim}.}
\label{fig:cut}
\end{figure}
\end{center}

\begin{center}
\begin{figure}
\epsscale{0.8}
\plotone{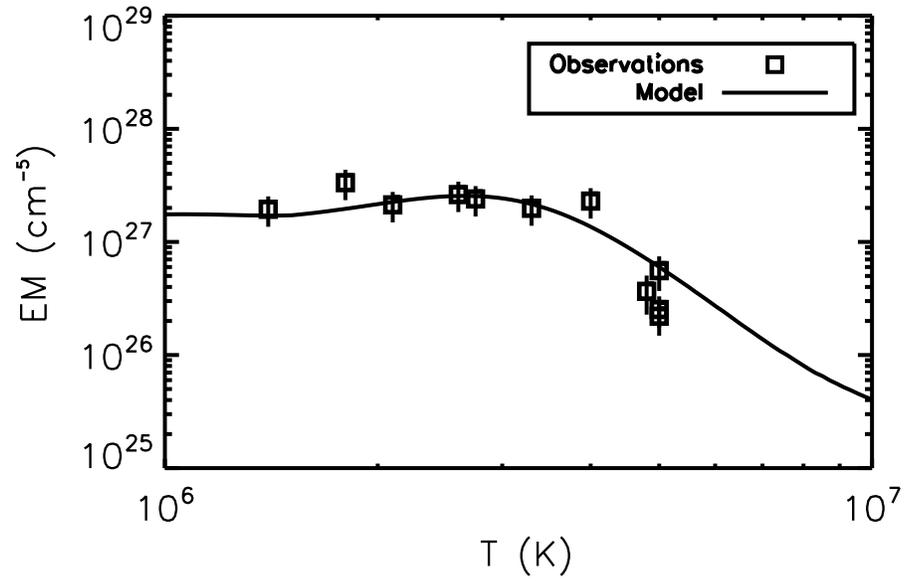}
\caption{Emission measure distribution of AR 10961.
Boxes: EIS observations; Solid line: Impulsively heated AR model.}
\label{fig:em}
\end{figure}
\end{center}

\begin{center}
\begin{figure}
\epsscale{0.8}
\plotone{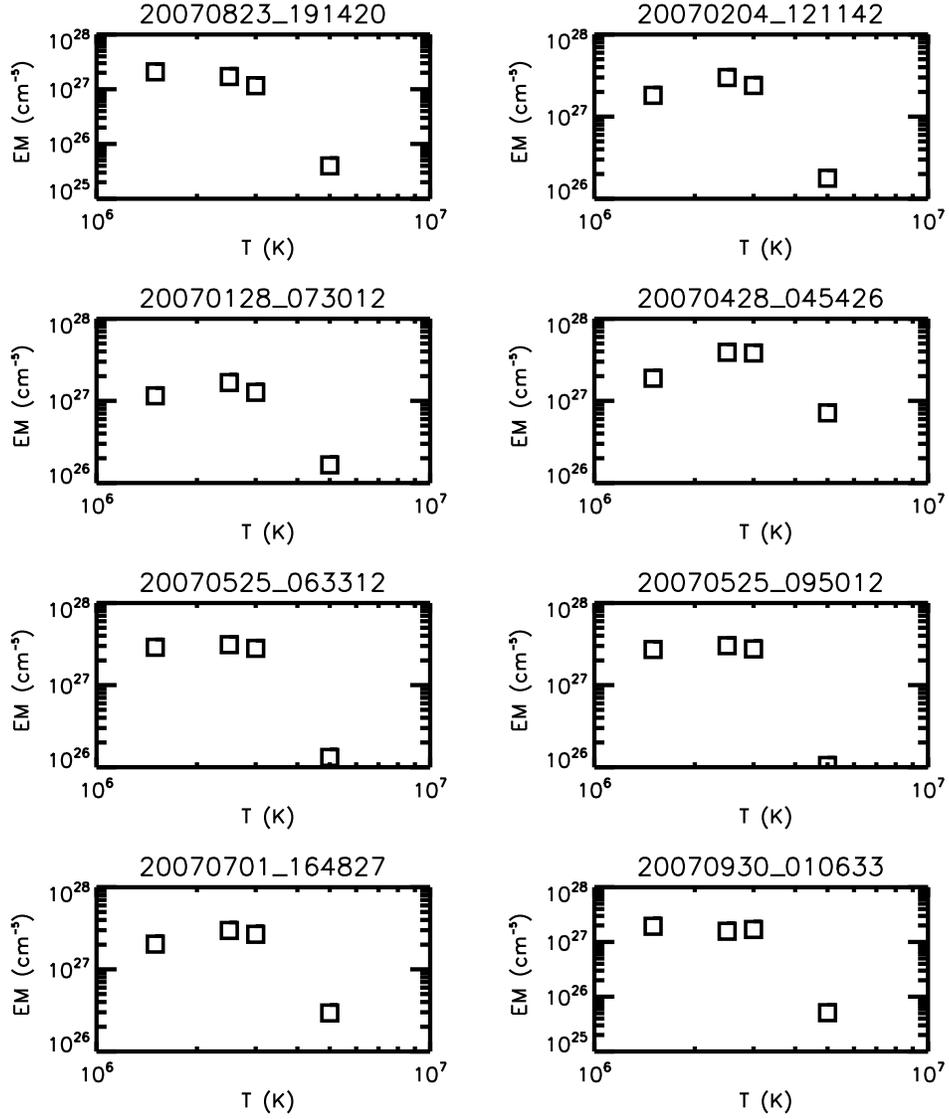}
\caption{Emission Measure distributions for 8 AR dataset observed in
the period January-September 2007. Intensities from Fe XII, Fe XV,
Ni XVII and Fe XVII (254.87 \AA\,) of Table 1 were used.}
\label{fig:ar_em}
\end{figure}
\end{center}

Knowing how distinct loops and the diffuse background contribute to
the total emission from an active region is an important question
that has been largely overlooked. Figure \ref{fig:cut} shows a
vertical intensity cut through the middle of the Fe XII image of
\ref{fig:arim}. There are a few local intensity maxima (e.g. around
solar-y 15, 30, 40) which can be identified with distinct loops.
However, these resolved loops do not stand out appreciably above the
background (they produce an intensity enhancement of only 10-30
$\%$) and they occupy a rather small fraction of the AR volume.
Therefore, we expect that the diffuse background contributes more to
our spatially averaged intensities and $EM$ distributions than do
distinct loops.  Of course part of the background could represent
indistinguishable, overlapping loops.  More work is needed on this
important question.

\section{Modeling}

A major goal of our study is to determine whether the observations
support or contradict impulsive coronal heating.   We therefore
calculated the $EM$ distribution expected from a simple model active
region that is heated by nanoflares.  We used our 0D hydrodynamic
simulation code called EBTEL, described in full by Klimchuk,
Patsourakos $\&$ Cargill (2008). For a given temporal profile of the
heating, EBTEL calculates the evolution of $EM(T)$ for both the
coronal and footpoint (i.e., transition region; TR) sections of a
stand. The inclusion of the TR emission in spatially averaged
intensities is important, since it can dominate the emission at
temperatures up to 1-2 MK or even higher, depending on the magnitude
of the nanoflare. EBTEL mimics complex 1D hydrodynamic simulations
very well (Klimchuk et al. 2008), but uses orders of magnitude less
computer time and memory. Note that impulsively heated AR models are
able to reproduce some of the salient morphological features of ARs
like the bright SXR core and the extended EUV loops (Warren $\&$
Winebarger 2007; Patsourakos $\&$ Klimchuk 2008).

We calculated hydrodynamic models for 26 stands with lengths in the
range 50-150 Mm, pertinent to the sizes of macroscopic loops in AR
10961. This model is a good approximation considering the  rather simple,
bipolar nature of the AR. We started with static equilibria having
an average coronal temperature near 0.5 MK. We heated the strands
with a triangular pulse lasting  50 s and let the strands cool for
8500 s, by which time the temperature had cooled below 1 MK.  The
amplitude $H$ of the heat pulses varied from strand to strand
according to
\begin{equation}
H=H_{0}{(L/L_{0})}^{\alpha},
\end{equation}
with $H_{0}$ the heating magnitude, $L_{0}$ the  length of the
shortest loop,  and $\alpha$ a scaling-law index that is related to
the specific mechanism of heating.  We chose $\alpha = -2.8$, which
is appropriate for heating that occurs when a critical shear angle
is reached in a magnetic field that is tangled by photospheric
convection (Mandrini et al. 2001; Dahlburg et al. 2005).

Following Klimchuk (2006) we time-averaged the $EM$ distributions
for each strand simulation to approximate a snapshot observation of
either a multi-stranded macroscopic coronal loop or an unstructured
background area. These individual distributions were then added
together to get a final $EM$ distribution for the entire active
region. Note that the heating magnitude $H_{0}$ of Equation 1 was
chosen to yield an $EM$ at warm ($\approx 1-2$) MK temperatures that
agrees with observed values. Previous studies attempting to
reproduce coronal loop observations in warm emissions adopted a
similar strategy. The open question has been whether the hot line
intensities predicted by the models are consistent with
observations. We here provide the first ever check of this type.

The solid curve in Figure \ref{fig:em} is the $EM$ distribution for
our model active region.  It agrees very well with the EIS
observations in the temperature range 1-5 MK for which we have data.
The model curve tracks both the warm plateau and the drop-off at
hotter temperatures. There is even a mild increase with temperature
below 3 MK, as seen in some of the examples of Figure
\ref{fig:ar_em}. This is the first time to our knowledge that an
impulsively heated AR model has been tested against spectroscopic
observations carried out over an extended temperature range, and
more importantly containing several hot lines, which supply the most
critical constraints to impulsive heating. The success of the model
provides further evidence that coronal heating is impulsive in
nature.

Our EIS dataset contained several very hot ($>10$ MK) ''flare" lines
(e.g. Fe XXIII, Fe XXV). Using the $EM$ distribution from our model
we found that the predicted intensities for these lines are too low
to be detected. We found no appreciable signal at the locations of
these lines in our EIS spectra, which serves as another test for our
model.

\section{Discussion and Conclusions}

We have argued that hot emission is an important diagnostic of
impulsive heating, and we have shown that the $EM$ distribution
predicted by a simple nanoflare-heated active region model agrees
well with distributions observed by EIS in the 1-5 MK temperature
range.  It is important to consider whether other heating scenarios
might be equally successful.  It seems likely that an appropriate
distribution of static equilibrium strands, corresponding to steady
heating, could also reproduce the observed distributions.  However,
such a model could not explain the over densities and other
properties of warm EUV loops that are fundamental constraints (e.g.,
Klimchuk 2006).

What if steady heating in hot loops were to suddenly shut off?  As
long as the loop begins at a high enough temperature, the cooling
plasma will be sufficiently over dense at 1-2 MK to explain the
observations. This does not seem to be a viable explanation,
however. Theory predicts that hot ($>$ 5 MK) static equilbrium loops
should very dense and therefore very bright. If warm EUV loops were
to result from the cooling of such loops, then we would expect to
observe an abundance of bright hot loops at the same locations as
warm loops. This is not the case. Hot loops are observed to be
fainter than expected for static equilibrium (Porter \& Klimchuk
1995). If they are monolithic structures (i.e., are fully filled
with hot plasma), then they are under dense.  If they have a small
filling factor, so as to be consistent with static equilibrium, then
they will not appear over dense when observed at warm temperatures
after the heating is shut off.  This scenario cannot explain the
observed over densities of warm EUV loops.

\begin{center}
\begin{figure}
\epsscale{0.8}
\plotone{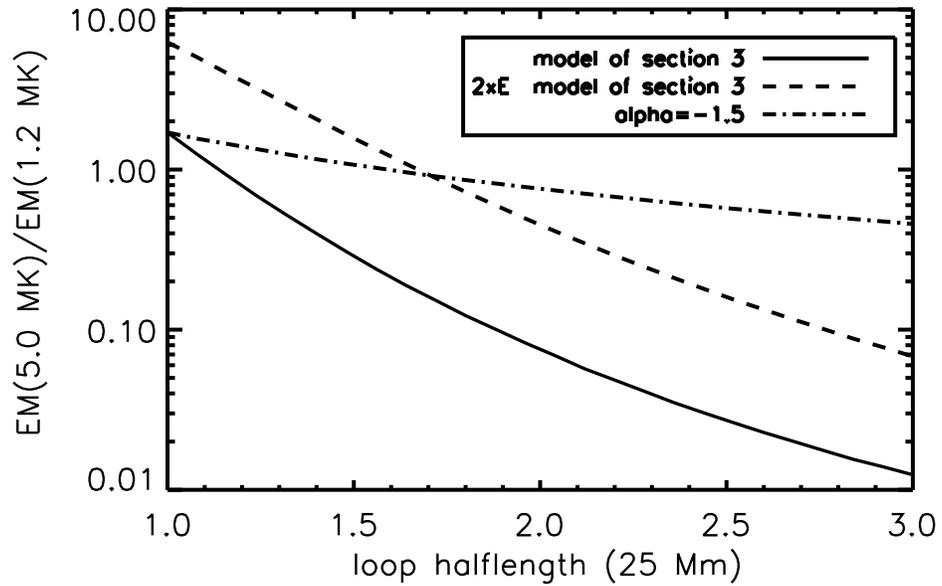}
\caption{Ratio of the temporally averaged Emission Measure at 5 and
1.2 MK as a function of strand halflength for every loop from AR
simulations: Section 2 (solid line), two-fold more energetic
nanoflares than these of Section 2 (dashes) and a shallower
dependance of nanoflare heating on strand length; the $\alpha$ of
Equation 1 is -1.5 (dashes-dots).}
\label{fig:em_hot_warm}
\end{figure}
\end{center}

We have seen that the hot emission predicted by impulsive heating is
faint compared to the warm emission.  We now show that the relative
intensity of the hot and warm emission has a dependence on loop
length that provides a useful diagnostic of nanoflare properties for
spatially resolved observations. The solid curve in Figure
\ref{fig:em_hot_warm} is the ratio of the temporally averaged
emission measure at 5 and 1.2 MK ($EM_{hot}/EM_{warm}$) as a
function of strand halflength for every strand from the AR
simulation of Section 2.  This ratio exhibits a very steep decrease
with strand length:  it falls off almost two orders of magnitude for
a two-fold increase in the strand length. Essentially, one would
expect hot loops and emissions to be seen within or close to AR
cores. However, if nanoflare heating has a weaker dependence on
strand length than assumed in Section 2 (i.e., if $\alpha > -2.8$ in
Equation 1), then $EM_{hot}/EM_{warm}$ varies much slower with $L$
(Figure 8, dot-dashed curve). Therefore, the chances of observing
hot emissions over extended areas will increase. The possibility of
detecting hot emissions in the outer parts of active regions is also
improved when the magnitude of the nanoflares $H_0$ increases
(dashed curve). We conclude that  plots of the radial distribution
of $EM_{hot}/EM_{warm}$  in ARs could serve as a diagnostic tool for
inferring  the properties of coronal heating.

Before closing we note that recent SXR and HXR broad-band and
spectroscopic observations by CORONAS-F, RHESSI and XRT have
demonstrated the existence of small yet measurable amounts of
emission at even higher temperatures ($\approx$ 5-12 MK) in
quiescent ARs (Zhitnik et al. 2006; McTiernan 2008; Siarkowski et
al. 2008; Reale et al. 2008). Moreover, a recent EIS study using the
hot Ca XVII line ($\approx$ 6 MK) showed loop emissions throughout
quiescent ARs (Ko et al. 2008). This line is heavily blended with a
Fe XI line and special care should be taken for subtracting off this
line from the Ca XVII line complex. These observations represent
further encouraging developments in the pursuit to identify the
coronal heating mechanism. Another important diagnostic of impulsive
heating is the development of wing asymmetries in the profiles of
hot lines such as Fe XVII (Patsourakos $\&$ Klimchuk 2007). We have
carried out detailed sit and stare observations in that line and
their analysis will be reported in the future.

Hinode is a Japanese mission developed and launched by ISAS/JAXA,
with NAOJ as domestic partner and NASA and STFC UK) as international
partners. It is operated by these agencies in cooperation with ESA
and NSC (Norway). This work was supported by the NASA Hinode and LWS
programs. We wish to thank H. Warren, I. Ugarte Urra, U. Feldman, C.
Brown,  E. Robbrecht, G. Doscheck and J. Mariska for helpful
discussions.

\clearpage

\begin{table}
\begin{center}
\caption{Spectral lines used in our analysis. $T_{form}$ is the formation temperature of
each line.}
\begin{tabular}{|c|c|c|}
\tableline\tableline
Line & Wavelength (\AA) & $T_{form}$ (MK) \\
\tableline
Fe XII &  195.12      & 1.4   \\
\tableline
Fe XIV  & 274.2 & 1.8 \\
\tableline
Fe XV &  284.16  & 2.1  \\
\tableline
Fe XVI & 262.98  &  2.6   \\
\tableline
Ni XVII & 249.18 & 2.8 \\
\tableline
Ca XIV & 193.87    &  3.3   \\
\tableline
Ca XV & 182.85 & 4.0  \\
\tableline 
Ca XVI & 208.6    &  4.8   \\
\tableline
Fe XVII & 204.67 & 5.0  \\
\tableline
Fe XVII &  254.87   &  5.0   \\
\tableline
Fe XVII & 280.16 & 5.0    \\
\tableline
\end{tabular}
\end{center}
\end{table}

\clearpage


\clearpage


\end{document}